# Semiconductor Solar Superabsorbers

Yiling Yu[2], Lujun Huang[1], Linyou Cao[1,2]*

[1]Department of Materials Science and Engineering, North Carolina State University, Raleigh NC 27695; [2]Department of Physics, North Carolina State University, Raleigh NC 27695;

**Abstract:**

Understanding the maximal enhancement of solar absorption in semiconductor materials by light trapping promises the development of affordable solar cells. However, the conventional Lambertian limit is only valid for idealized material systems with weak absorption, and cannot hold for the typical semiconductor materials used in solar cells due to the substantial absorption of these materials. Herein we theoretically demonstrate the maximal solar absorption enhancement for semiconductor materials and elucidate the general design principle for light trapping structures to approach the theoretical maximum. By following the principles, we design a practical light trapping structure that can enable an ultrathin layer of semiconductor materials, for instance, 10 nm thick a-Si, absorb > 90% sunlight above the bandgap. The design has active materials with one order of magnitude less volume than any of the existing solar light trapping designs in literature. This work points towards the development of ultimate solar light trapping techniques.

* To whom correspondence should be addressed.

Email: lcao2@ncsu.edu

Maximizing the enhancement of solar absorption in semiconductor materials by light trapping promises the development of extremely cost-effective solar cells.[1-7] A maximized enhancement can enable the full absorption of incident solar radiation with a minimal volume of semiconductor materials. This may lead to a dramatic reduction in the cost of materials and material processing for the manufacturing of solar cells, which is widely considered as a key step towards substantially lowering the overall cost of solar cells. However, *what is the maximal enhancement theoretically attainable by light trapping for semiconductor materials* and *how to design light trapping structures to realize the maximal enhancement in practical devices* have essentially remained unanswered yet. Previous studies on the upper limit of solar absorption enhancement are usually limited to idealized materials that are assumed to have very weak absorption.[1, 7-10] The assumption of weak absorption is necessary for exploring the statistic approaches of these studies to find out the limit. In contrast, the semiconductor materials used in practical solar cells, such as amorphous silicon (a-Si), CdTe, and CIGS all have substantial absorption, far beyond the weak absorption assumed for the idealized materials. As a result, the rational of the previous studies and the upper limit derived thereby, for instance, the Lambertian limit, cannot hold for the typical semiconductor materials in solar cells.

It is very challenging to find out the maximal solar absorption enhancement and associated design principles for semiconductor materials. The absorption of semiconductor materials strongly depends on the physical features, including shape, surface texture, and dimensionality, of the materials.[11-24] To find out the maximal solar absorption enhancement would request evaluation and comparison of the solar absorption as a function of all the possible physical features. For instance, to find out the maximal solar absorption in a semiconductor material with a given volume, we would have to examine the solar absorption of the material in forms of all kinds of shapes, sizes, and structures, which would amount to an infinite number of varieties. None of the current methods is able to perform such a thorough analysis. Typical numerical or analytical methods, such as Mie theory, transfer matrix method (TMM), finite difference time domain (FDTD), can precisely evaluate the solar absorption of semiconductor materials.[25, 26] But these methods rely on rigorously matching boundary conditions at the surface of the materials and involve intensive computation efforts. Every change in the physical features would request a new calculation in order to find out the solar absorption. The infinite variety of physical features simply makes it impossible to search for the maximal solar absorption enhancement using these methods.

Here we demonstrate the maximal solar absorption enhancement theoretically attainable by light trapping for semiconductor materials and elucidate the principles for the rational design of light trapping structures to approach the maximum. By following the principles, we present a general design of light trapping structures that can enable an ultra-thin semiconductor materials to absorb >90% incident solar light above the bandgap. These include 10 nm thick a-Si or 50 nm thick CdTe or 30 nm thick CIGS. This design has active materials with one order of magnitude less volume than any of the sophisticated solar light trapping designs in literature. It is close to the theoretically predicted maximal solar absorption enhancement (solar superabsorption) and indeed shows enhancement beyond the conventional Lambertian limit.

This study leverages on an intuitive model, coupled leaky mode theory (CLMT), that we have recently developed for the analysis of light absorption in semiconductor structures.[23, 27, 28] The CLMT model considers a semiconductor structure as a leaky resonator and analyzes the light absorption of the structure as a result of the coupling between incident light and the structure's leaky modes (Figure 1a inset). The key advantage of the CLMT model is transforming the variables for the analysis of solar absorption. In contrast with most of the existing approaches, which use physical feature as variables (shape, surface texture, dimensionality, etc.) in the analysis of solar absorption, the CLMT model can evaluate solar absorption with only two leaky mode variables, radiative loss and resonant wavelength, regardless the physical features of the materials. This variable transformation provides a unique capability to find out the maximal solar absorption enhancement of a material by surveying over the two leaky mode variables, instead of searching among the physic features with infinite number of varities. The survey can also specify the requirements for the two variables, for instance, the optimal value of the variables, in order to achieve the maximal solar absorption enhancement. This knowledge can serve as a very useful guide for the rational design of light trapping structures. Essentially, in order to achieve the maximal solar absorption enhancement, what we need do is to design light trapping structures whose leaky modes can satisfy the requirements on radiative loss and resonant wavelength.

Note that similar mode coupling approaches have been used for the analysis of light absorption previously.[1, 29-31] However, unlike all the previous mode-coupling works that focus on materials with weak intrinsic absorption and confined optical modes, our studies extend the mode coupling to strong absorbing materials and to leaky optical modes with small quality factors, which are of particular importance for applications in solar cells. Another novel discovery of our studies is the transformation of the variables, i.e. correlating solar absorption to two leaky mode variables instead of physical feature variables that may amount to an infinite number. These two breakthroughs lay down the groundwork for us to search the maximal solar absorption enhancement in semiconductor materials, which has been long believed very difficult.

The coupled leaky mode theory (CLMT) model considers the light absorption of a semiconductor structure as result of the coupling between incident light and the structure's leaky modes (Fig. 1a). Leaky modes are defined as natural optical modes with propagating waves outside the structure. Each of the leaky modes is featured with a complex eigenvalue ($N_{real}$ - $N_{imag}.i$), which can be analytically or numerically solved out.[22, 27, 28] For materials with a refractive index of $n$ ($n = n_{real} + n_{imag}.i$), the absorption cross-section $C_{abs}$ contributed by one leaky mode can be derived from[27, 28]

$$C_{abs} = F \frac{2 N_{imag}/N_{real} \cdot n_{imag}/n_{real}}{(\alpha-1)^2 + (N_{imag}/N_{real} + n_{imag}/n_{real})^2} \cdot Corr \qquad (1)$$

where $\alpha$ indicates the offset between the incident wavelength $\lambda$ and the resonant wavelength of the leaky mode $\lambda_0$ as $\alpha = n_\lambda \lambda_0 / n_0 \lambda$, $n_\lambda$ and $n_0$ are the real part of the refractive index of the materials at $\lambda$ and $\lambda_0$, respectively. The resonant wavelength $\lambda_0$ is related with the real part of the eigenvalue as $\lambda_0 = 2\pi n_0 r / N_{real}$, where $r$ is the characteristic size of the structure, for example, the radius of nanoparticles (NPs). $F$ is a factor mainly associated with the dimensionality of the materials. It can be found as $\lambda/\pi$ and $\lambda^2/4\pi$ for 1D circular nanowires (NWs) and 0D spherical NPs, respectively.[27, 28] These values can also be reasonably applied to structures with other shapes, such as rectangle and triangle.[23, 27] $Corr$ is a correction term that limits the leaky mode to only couple incident wavelengths at the proximity of its resonant wavelength. It does not have a rigorous expression, but can be found using numerical fitting. We find that $1/[1+4(\alpha-1)^2]$ for $\alpha > 1$ or $1/[1+4(1/\alpha-1)^2]$ for $\alpha < 1$ is a reasonable approximation.[23, 27] The solar absorption cross-section $P_{solar}$ of the leaky mode can be calculated by integrating $C_{abs}$ over the spectral flux of solar radiation $I_\lambda$ as

$$P_{solar} = \int_\lambda I_\lambda C_{abs} \, d\lambda \qquad (2)$$

For structures with multiple leaky modes, the total absorption is just a simple sum of the absorption contributed by each of the modes. For the convenience of discussion, we assume that every single solar photon absorbed can be converted into one electron. Therefore, the calculated solar absorption in this work has a unit similar to that of short-circuit currents.

We can confirm the validity of eqs. (1)-(2) by comparing the results with those obtained with well-established models. We use 0D a-Si nanoparticles (NPs) as an example. Fig. 1b-c shows the spectral absorption cross-section $C_{abs}$ and solar absorption $P_{solar}$ of single a-Si NPs calculated using eqs. (1)-(2) and the well-established Mie theory.[25] The eigenvalues of leaky modes in the a-Si NP can be analytically solved as we demonstrated previously.[27, 28] The refractive index of a-Si is obtained from references.[32] The calculation results of eqs. (1)- (2) show reasonable agreement with those of the Mie theory. Eqs. (1)-(2) can be confirmed generally valid for semiconductor structures with other materials, shapes, and dimensionalities (Figure S1). It is worthwhile to note that the CLMT is in essence an approximate model. However, the compromise in accuracy is well compensated by its intuitiveness and simplicity that cannot be obtained from other methods.

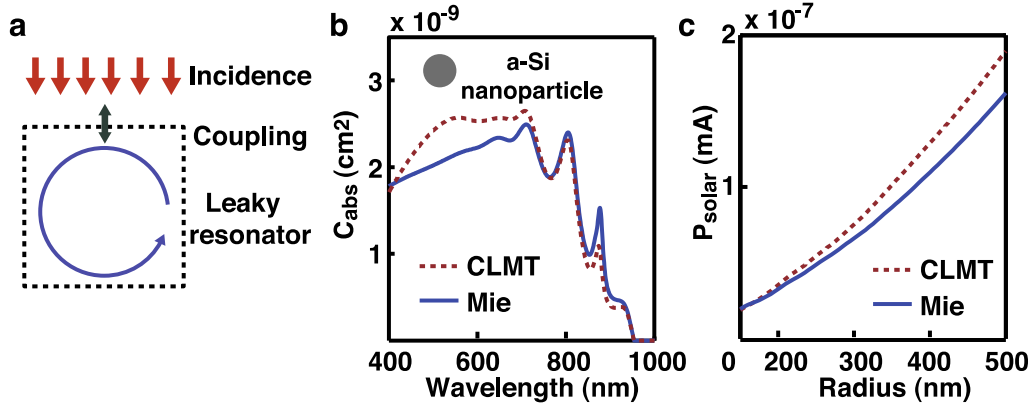

**Figure 1. Comparison of the coupled leaky mode theory (CLMT) with the well-established Mie theory.** (a) Schematic illustration of the coupling of incident light with the leaky mode of an arbitrary structure. The dashed line indicates the structure could have any arbitrary shapes. (b) The spectral absorption efficiency of a 250 nm radius a-Si nanoparticle calculated using the Mie theory (solid blue line) and the CLMT model (dash red line). The inset is a schematic illustration of the nanoparticle. (b) The solar absorption efficiency of single a-Si NPs as a function of the radius calculated using the Mie theory (solid blue line) and the CLMT model (dash red line).

Eqs. (1)-(2) provides a variable transformation that is useful for the search of the maximal solar absorption. It demonstrates that the solar absorption can be equivalently evaluated using leaky mode variables rather than physical feature variables, such as shape, size, surface texture, and dimensionality, as done by most of the existing methods. The equations indicate that the solar absorption of a material with known refractive index $n$ is dictated by only two leaky mode variables, the radiative loss $q'_{rad}$ ($q'_{rad} = N_{imag} / N_{real}$) and the resonant wavelength $\lambda_0$, regardless the physical features of the material. Assuming only one leaky mode exists in the material, the maximal solar absorption of the material can be found out by evaluating eq.(2) as a function of the two leaky mode variables. This calculation is also expected to elucidate the value of the two variables associated with the maximal solar absorption.

Fig. 2a shows the solar absorption of single-mode 0D a-Si structures calculated using eq. (2) as functions of the radiative loss $q'_{rad}$ and resonant wavelength $\lambda_0$. We can immediately find that the leaky mode with resonant wavelength in the range of 600 - 650 nm and radiative loss in the range of 0.12 - 0.22 may have the maximal solar absorption of $1.88 \times 10^{-9}$ mA. This suggests that to maximize the solar absorption in a single-mode a-Si structure requests the structure to be designed such to have the leaky mode's radiative loss and resonant wavelength lying in these optimal ranges, respectively. Similar optimal ranges of the two variables can also be found with other semiconductor materials (Fig. S2-S3). Generally, the optimal range of the resonant wavelength is to match the photon flux of the solar spectrum, while the optimal range of the radiative loss is to reasonably match the intrinsic absorption loss ($n_{imag}/n_{real}$) of the materials, which can create a desired "critical coupling" between incident light and the leaky mode.[27, 29, 33]

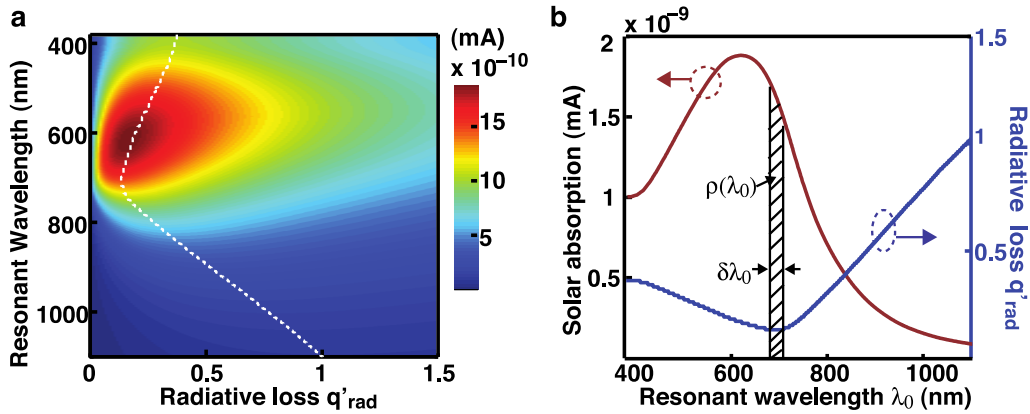

**Figure 2. The solar absorption of single leaky modes in 0D a-Si structures.** (a) Calculated solar absorption of single leaky modes in 0D a-Si structures as a function of radiative loss (horizontal axis) and resonant wavelength (vertical axis). The white dashed line connects the optimal absorption at each resonant wavelength. (b) The optimal absorption (red line) and associated radiative loss (blue line) as a function of the resonant wavelength. The shaded area is to schematically illustrate the integration of the contributions from multiple leaky modes.

The result for single leaky modes provides the capability to find out the maximal solar absorption of multi-mode structures that are typically used in solar cells. The multiple leaky modes would have different resonant wavelengths. To maximize the solar absorption in a multi-mode structure requests the absorption of each mode to be optimized. We can identify the optimal solar absorption of a leaky mode with an arbitrary resonant wavelength, as illustrated by the white line in Fig. 2a, and replot the optimal solar absorption and the associated radiative loss as a function of the resonant wavelength $\lambda_0$ in Fig. 2b. The maxima solar absorption of the multiple-mode structure can be derived from

$$P_{max} = \int_{\lambda_0} P_{opt}(\lambda_0)\rho(\lambda_0)d\lambda_0 \qquad (3)$$

$P_{opt}(\lambda_0)$ and $\rho(\lambda_0)$ are the optimal solar absorption of one leaky mode and the density of leaky modes at an arbitrary resonant wavelength $\lambda_0$. We find that the density of leaky modes follows the well-established formalism of mode density in optical resonators (see S3 of the Supplementary Information).[34] $\rho(\lambda_0) = 8\pi n_0^3 V/\lambda_0^4$ for 0D structures with arbitrary shapes, where $V$ is the volume of the structure (see Figure S4-S5). This expression can also be reasonably applied to heterostructures that consist of both absorbing and non-absorbing materials, for instance, core-shell structures, in which $V$ is the volume of the absorbing materials (see Figure S6-S7).

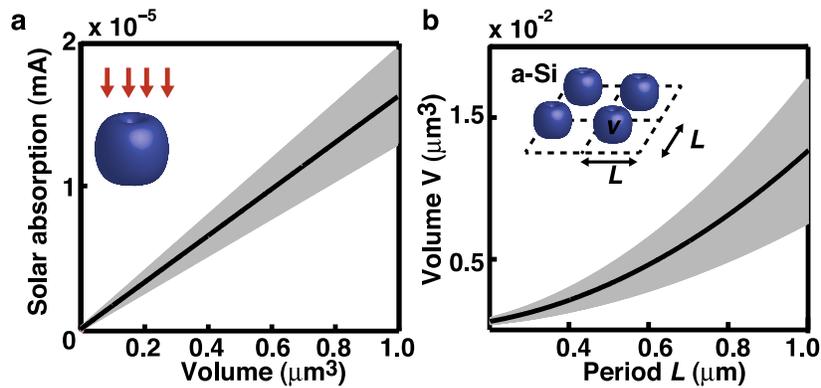

**Figure 3. Solar superabsorption in single nanostructures and an array of nanostructures.** (a) The maximal solar absorption of single 0D nanostructures ( a-Si is included as the absorbing materials) as a function of the volume of a-Si materials. The calculation result includes an estimated 20% error as indicated by the shaded areas. The inset is a schematic illustration for the nanostructure, whose irregular shape is intentionally chosen to illustrate that the structure may have any arbitrary shapes. (b) Solar superabsorption limit of an array of 0D nanostructures. The minimum volume of a-Si materials necessary to absorb >90% of the solar radiation above the band gap is plotted as a function of the period of the array. The inset is a schematic illustration for the nanostructure array.

Fig. 3a plots the maximal solar absorption of a multi-mode 0D structure obtained from the evaluation of eq. (3) on the basis of the result given in Fig.2b. Again, a-Si is used as the absorbing materials in the structure. The absorption maximum shows a linear dependence on the volume $V$ of a-Si materials in the structure. This is due to the linear dependence of the mode density on the volume, which indicates that the number of leaky modes fundamentally limits the amplitude of the maximal solar absorption. It is worthwhile to note that 0D structures are typically more favorable for applications in solar absorption than 1D structures due to a higher density of leaky modes. The result in Fig. 3a involves an estimated error of 20%, as indicated by the shaded area. The error mainly results from the approximate nature of the CMLT model and a finite spectral integration in the evaluation of eq. (3). Eq.(3) would be ideally integrated over the entire spectrum of resonant wavelengths from 0 to infinity, but in reality we can only perform the integration in a finite range (300-1700nm used for Fig. 3a), which is limited by the refractive index of the absorbing materials available in references.[32]

From the perspective of practical application, of the most interest is to find out the minimal volume of absorbing materials in a large-scale structure, for instance, an array of nanostructures, to absorb most (> 90%) of the incident solar light above the band gap. This is now possible with the knowledge of the maximal solar absorption in single structures (Fig. 3a). Without losing generality, we use a square periodic array of 0D nanostructures as an example (Fig. 3b inset). We assume that the light absorption in each individual structure of the array is similar to that of one single structure. Our previous studies have already demonstrated that this is a reasonable assumption.[17, 24] Therefore, what we need is to find out the minimum volume of a-Si materials in each individual structure necessary to absorb 90% of the incident solar radiation inside one unit space as 90%×22.7× $L^2$ (22.7 mA/cm$^2$ is all the solar energy above the bandgap of a-Si). We can readily find out the minimum volume in Fig. 3a and plot it as a function of $L$ in Fig. 3b. This result involves an estimated error of 40%, in which 20% is inherited from the calculation for single structures and another 20% from the possible difference in the solar absorption of one individual nanostructure in the array from that of one single nanostructure. Similar analysis for the maximal solar absorption in single nanostructures and nanostructure arrays of other semiconductor materials, such as CdTe, can be seen in Fig. S8 in the Supporting Information.

The CLMT analysis may also suggest the general principles for the rational design of structures to approach the predicted maximal solar absorption. As illustrated in Fig. 2, to maximize the solar absorption requests the radiative loss of leaky modes to be in an appropriate range that roughly match the intrinsic absorption loss ($n_{imag}/n_{real}$) of the absorbing materials. However, the radiative loss of typical semiconductor structures tends to quickly decrease with the mode number increasing. For instance, the radiative loss of the leaky mode in 0D a-Si structures would ideally be in the range of 0.05-0.7 in order to have substantial solar absorption. But in typical 0D semiconductor structures only one leaky mode can satisfy this requirement.[27, 28] The limited number of the leaky modes with appropriate radiative loss is the reason why the solar absorptions of typical semiconductor nanostructures reported are far less than the predicted solar absorption maximum shown in Fig. 3.[17] Therefore, we can conclude that key to maximize the solar absorption is to engineer the modes in semiconductor structures to be more leaky (larger radiative loss).

Heterogeneous structures, such as core-multishell nanostructures that consists of absorbing and non-absorbing materials, provide a promising platform to engineer the radiative loss of leaky modes.[24] While non-absorbing materials cannot absorb solar light by themselves, heterostructuring these materials, which typically have lower refractive indexes, with absorbing semiconductor materials can change the dielectric environment of the semiconductor materials, which may subsequently lead to increase in the radiative loss of leaky modes and hence increase in the solar absorption. There are two basic heterostructuring strategies, coating absorbing materials with non-absorbing materials (absorbing core/non-absorbing shell) or replacing part of absorbing materials with non-absorbing materials (non-absorbing core/absorbing shell) (Fig. 4a). The two strategies can be used together (non-absorbing core/absorbing/non-absorbing shell) (Fig. 4a). The physics underlying the leaky mode engineering with the heterostructures can be intuitively understood from the perspectives of impedance match and effective refractive index. The radiative loss physically indicates how easily the structure can couple (radiate) light out to its environment. According to the reciprocal nature of light, a larger radiative loss would also mean an easier coupling of incident solar radiation into the structure. As illustrated in Fig. 4a, the high refractive index contrast of semiconductor structures with environment poses the major barrier for the in-coupling of incident solar light. A coating with gradually changed refractive index can help minimize the impedance mismatch and facilitate the coupling of incident light into the structure. Additionally, replacing part of the high-index semiconductor materials with low-index, non-absorbing materials may lower the effective refractive index of the structure, which can also facilitate the in-coupling.

We can use core-multishell nanowires (NWs) as an example to further illustrate the notion of engineering the radiative loss of leaky modes. The nanowire is used here because it can provide a better visualization of the eigenfield of leaky modes than a nanoparticle. The principles developed with the NW can also apply to nanoparticles. Fig. 4b-d shows the calculated eigenfield of one leaky mode in three different structures, a solid semiconducor NW in radius of 140 nm (structure 1, Fig. 4b), a NW consisted of a 130nm radius dielectric core and a10nm thick semiconductor coating (structure 2, Fig.4c), and a NW consisted of a 130 nm radius core, a10nm thick semiconductor coating, and other three shell layers in thickness of 60nm, 50nm, and 60nm from the inner to outer, respectively (structure 3, Fig.4c). The three shell layers are non-absorbing with refractive index being 2.7, 2.0, and 1.5 from the inner to outer. The refractive indexes of the semiconductor materials and the non-absorbing core are set to be a constant of 4 (close to that of a-Si) and 2 (close to that of ZnO), respectively. We can see that the eigenfield turns to be much more spread in the heterostructures with the radiative loss calculated to be 0.002, 0.036, and 0.115 for the structure 1, 2, and 3, respectively. Similar substantial increase in the radiative loss can also be seen in other leaky modes ( Table S1). As a result, the structure 3, with a much less (13.7%) amount of absorbing materials, shows a absorption more than twice as big as the solid NW.

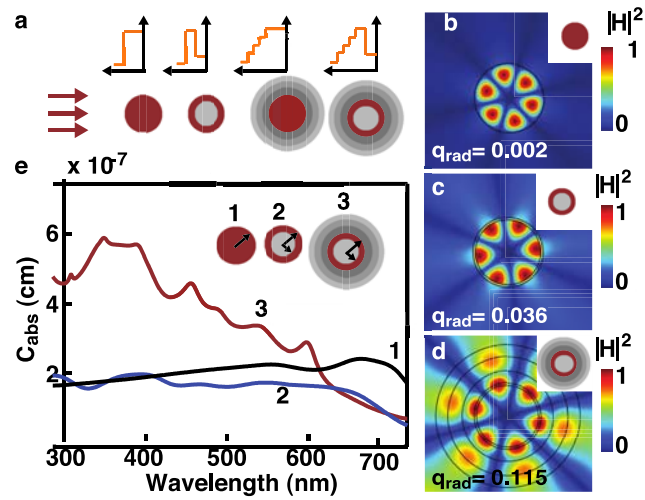

**Figure 4. Leaky mode engineering in heterostructures.** (a) Schematic illustration for the refractive index profile in homogeneous and various heterogeneous structures. (b-d) The eigen magnetic field distribution of one leaky mode ($TE_{31}$) in a solid semiconductor NW in radius of 140 nm, a core-shell NW consisted of a 130nm radius dielectric core and a 10nm thick semiconductor coating, and a core-multishell NW consisted of a 130 nm radius core, a 10nm thick semiconductor layer, and other three shell layers in thickness of 60nm, 50nm, and 60nm from the inner to outer, respectively. The refractive indexes of the semiconductor and the dielectric core are set to be 4 and 2, respectively. Those of the three shell layers are set to be 2.7, 2.0, and 1.5 from the inner to outer. (e) Calculated spectral solar absorption for the three structures with the semiconductor materials being a-Si.

Using the strategy of engineering leaky modes with core-multishell structures, we can design arrays of 0D nanostructures whose solar absorption can indeed approach the predicted solar superabsorption as shown in Fig. 3b. Fig.5a-b shows an example of the design. This design is a square array of quasi-core-shell structures on top of silicon oxide substrates. The structures are designed to be in a rectangular shape for the convenience of fabrication. Each individual structure consists of a non-absorbing ZnO core in size of (180 ± 10) × (180± 10) × (380 ± 10) nm that is conformally coated by a-Si (15± 5 nm), SiC (30 ± 10 nm), ZnO (30 ± 10 nm), and $SiO_2$ (50 ± 10 nm). The period of the array is 540 ± 60 nm. The given numerical error the tolerance of this design in geometrical features (Fig. S9), which suggests that it is robust for manufacturing. SiC, ZnO, and $SiO_2$, with refractive indexes of 2.7, 2.0, and 1.5, respectively, are chosen to provide a gradually changed refractive index in the coating to minimize the impedance mismatch. The substrate can essentially be in any arbitrary thickness and is set to be in thickness of 50 μm in this work. A mirror is designed underneath the substrate. Note that the presence of the mirror may not change the radiative loss of leaky modes, but can affect the coupling of incident solar light with the nanostructure by alternating coupling channels. The nanostructure array may electromagnetically couple with environment through the channels at both sides of the array. The mirror may turn off all the channels at one side, which may facilitate the coupling of the nanostructure with the light incident from the other side. Intuitively, without the mirror, the incident light could more likely pass through the nanostructure array without being absorbed.

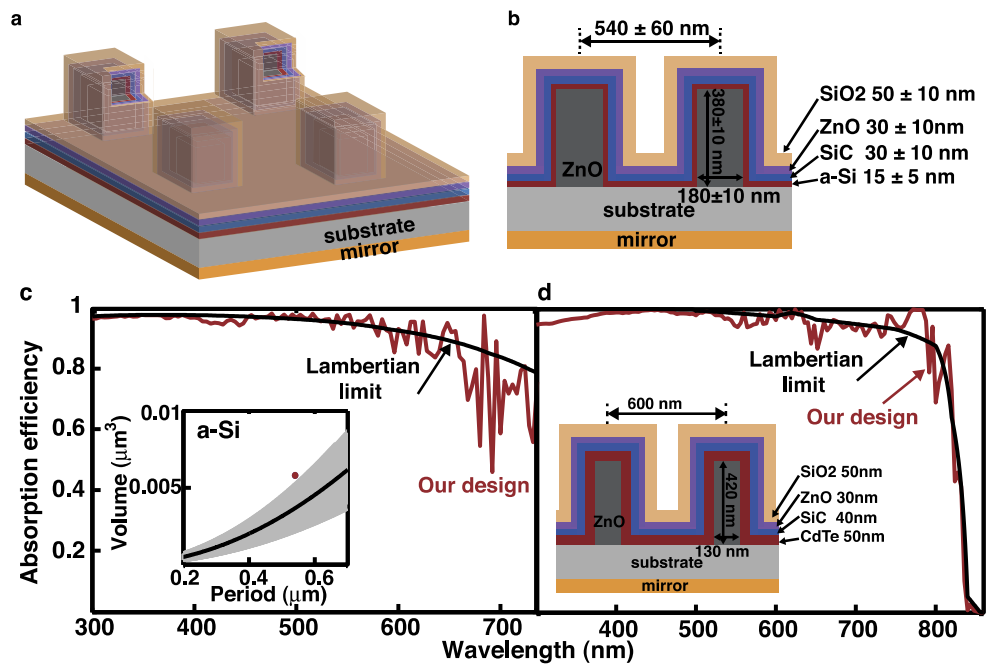

**Figure 5. Design of solar superabsorbers.** (a) Schematic illustration for the designed nanostructure array. (b) Geometrical features of the structure designed for a-Si solar superabsorbers. (c) The calculated spectral absorption efficiency of the structure shown in (b), also given in the Lambertian limit for a-Si with an efficient thickness of 19.4 nm. The inset shows the relationship between the designed structure (the red dot) and the predicted minimum volume. (d) The calculated spectral absorption efficiency of a designed structure including 50nm thick CdTe. The Lambertian limit for CdTe with an efficient thickness of 80 nm is also given (black). The inset shows the geometry of the designed structure.

This design can efficiently absorb solar light. We can see that a layer of a-Si in thickness of 10 nm can absorb 91% of all the solar radiation above the bandgap (734 nm), which amounts to 20.7mA/cm$^2$. In many ranges of the solar spectrum the absorption of the designed structure is even better than the Lambertian limit,[4, 7] $4n_{real}^2\alpha d /(1+4n_{real}^2\alpha d)$, where $\alpha$ is the absorption coefficient of a-Si materials and $d$ is the effective thickness of 19.4 nm ( the effective thickness is calculated by assuming the same volume of materials in a format of planar films). This confirm that the Lambertian limit derived from idealized materials indeed cannot be applied to real materials. The volume of a-Si materials in each nanostructure can be calculated as 0.0058 μm$^3$. Although this number is not as small as the predicted minimum volume as shown in Fig. 3b, it is nevertheless very close (Fig. 5c inset). The design can be reasonably applied to other materials, such as CdTe and CIGS. By using a similar design (Fig. 5d inset), we can enable a layer of 50 nm CdTe to absorb 90% of all the solar radiation above the bandgap as 27.7mA/cm$^2$. Again, the absorption of the designed structure can be seen better than the Lambertian limit in many spectral ranges (Fig. 5b) and reasonably close to the predicted minimum volume for CdTe materials (Fig. S10). We can also find that a layer of 30 nm thick CIGS in another similar design can be enabled to absorb 90% of the solar radiation above the bandgap (Fig. S11).

It should be noted that our designs by no means the best design in theory. We can find that, although very close, the designed structure need more volume of absorbing materials than the prediction for the targeted solar absorption (Fig. 5c inset, Fig. S10, and Fig. S11b). One key parameter in the design is the refractive index profile of the multilayer non-absorbing shell. The ideal design would have a smoothly changed refractive index from environment to the absorbing materials. However, in reality this would be limited by the lack of non-absorbing materials with arbitray refractive index, in particular, high index close to that of the semiconductor materials. It would also be limited by possible manufacturing challenges in produce such a coating with nanoscale meters. Our designs involve only three layers of SiC, ZnO, and $SiO_2$, which can be manufactured in cost-effective ways. The excellent absorption of our designed structures provides very strong support for the maximal solar abosprtion and the design principles predicted by our CLMT analysis. It should also be noted that our designs by no means the only one that can approach the maximal solar absorption. We believe that there are many other structures that can achieve similar or even better solar absorption. But regardless whatever structures are designed, the radiative loss of the leaky modes involved would have to be engineered into the same range.

## Acknowledgements

This work is supported by start-up fund from North Carolina State University. L. C. acknowledges a Ralph E. Power Junior Faculty Enhancement Award from Oak Ridge Associated Universities.

## Author Contributions

Y.Y. and L.C. conceived the idea, developed the analytical model, performed the modeling and data analysis. L.H. contributed to the numerical calculation. All authors were involved in writing the paper.

## Additional Information

Competing financial interests: The authors declare no competing financial interests.

**Figure Legends**

**Figure 1. Comparison of the coupled leaky mode theory (CLMT) with the well-established Mie theory.** (a) Schematic illustration of the coupling of incident light with the leaky mode of an arbitrary structure. The dashed line indicates the structure could have any arbitrary shapes. (b) The spectral absorption efficiency of a 200 nm radius a-Si nanoparticle calculated using the Mie theory (solid blue line) and the CLMT model (dash red line). The inset is a schematic illustration of the nanoparticle. (b) The solar absorption efficiency of single a-Si NPs as a function of the radius calculated using the Mie theory (solid blue line) and the CLMT model (dash red line).

**Figure 2. The solar absorption of single leaky modes in 0D a-Si structures.** (a) Calculated solar absorption of single leaky modes in 0D a-Si structures as a function of radiative loss (horizontal axis) and resonant wavelength (vertical axis). The white dashed line connects the optimal absorption at each resonant wavelength. (b) The optimal absorption (red line) and associated radiative loss (blue line) as a function of the resonant wavelength. The shaded area is to schematically illustrate the integration of the contributions from multiple leaky modes.

**Figure 3. Solar superabsorption in single nanostructures and an array of nanostructures.** (a) The maximal solar absorption of single 0D nanostructures ( a-Si is included as the absorbing materials) as a function of the volume of a-Si materials. The calculation result includes an estimated 20% error as indicated by the shaded areas. The inset is a schematic illustration for the nanostructure, whose irregular shape is intentionally chosen to illustrate that the structure may have any arbitrary shapes. (b) Solar superabsorption limit of an array of 0D nanostructures. The minimum volume of a-Si materials necessary to absorb >90% of the solar radiation above the band gap is plotted as a function of the period of the array. The inset is a schematic illustration for the nanostructure array.

**Figure 4. Leaky mode engineering in heterostructures.** (a) Schematic illustration for the refractive index profile in homogeneous and various heterogeneous structures. (b-d) The eigen magnetic field distribution of one leaky mode ($TE_{31}$) in a solid semiconductor NW in radius of 140 nm, a core-shell NW consisted of a 130nm radius dielectric core and a10nm thick semiconductor coating, and a core-multishell NW consisted of a 130 nm radius core, a10nm thick semiconductor layer, and other three shell layers in thickness of 60nm, 50nm, and 60nm from the inner to outer, respectively. The refractive indexes of the semiconductor and the dielectric core are set to be 4 and 2, respectively. Those of the three shell layers are set to be 2.7, 2.0, and 1,5 from the inner to outer. (e) Calculated spectral solar absorption for the three structures with the semiconductor materials being a-Si.

**Figure 5. Design of solar superabsorbers.** (a) Schematic illustration for the designed nanostructure array. (b) Geometrical features of the structure designed for a-Si solar superabsorbers. (c) The calculated spectral absorption efficiency of the structure shown in (b), also given in the Lambertian limit for a-Si with an efficient thickness of 19.4 nm. The inset shows the relationship between the designed structure (the red dot) and the predicted minimum volume. (d) The calculated spectral absorption efficiency of a designed structure including 50nm thick CdTe. The Lambertian limit for CdTe with an efficient thickness of 80 nm is also given (black). The inset shows the geometry of the designed structure.

*Supplementary Information*

*for*

# Semiconductor Solar Superabsorbers

Yiling Yu[2], Lujun Huang[1], Linyou Cao[1,2]*

[1]Department of Materials Science and Engineering, North Carolina State University, Raleigh NC 27695;
[2]Department of Physics, North Carolina State University, Raleigh NC 27695;

* To whom correspondence should be addressed.

Email: lcao2@ncsu.edu

## This PDF file includes

Fig. S1-Fig. S3

S1. Calculation for the density of leaky modes including Fig. S4-S7

Fig. S8-Fig.S11

Table S1

Reference S1-S3

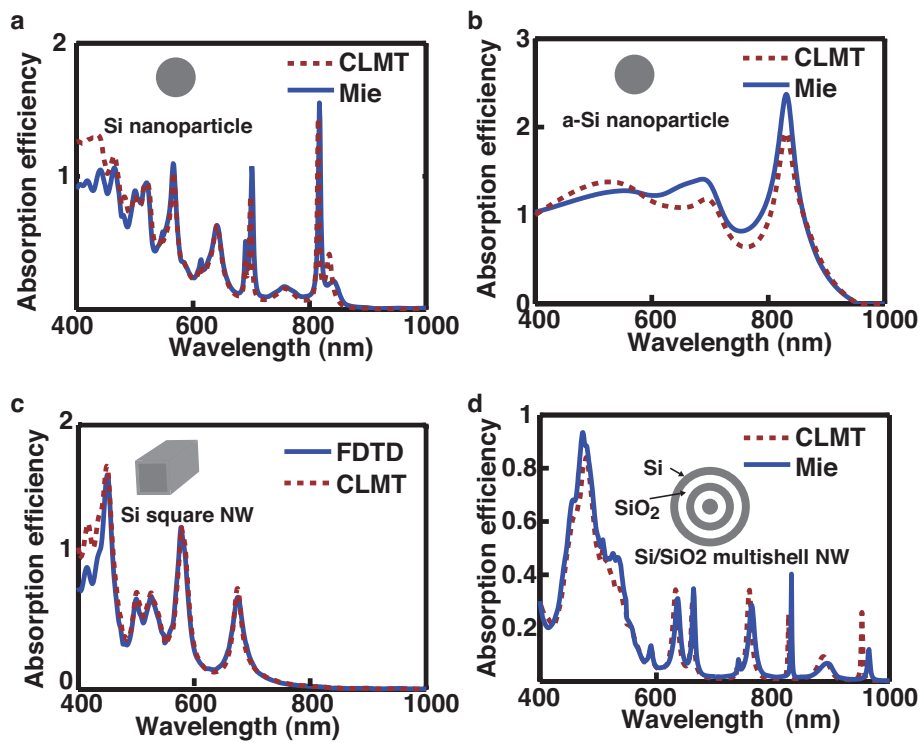

Figure S1. The general validity of CLMT for the analysis of light absorption in semiconductor nanostructures. Calculated spectral light absorption efficiency using the CLMT model (red dashed lines) and well-established analytical or numerical methods (blue solid lines) for (a) Si nanoparticles in radius of 150 nm, (b) a-Si nanoparticles in radius of 150 nm, (c) Si square nanowires in size of 200 nm, and (d) core-multishell Si/SiO2 nanowires in radius of 300 nm, which equally distributes among the five layers involved. Insets are schematic illustrations for the nanostructures calculated.

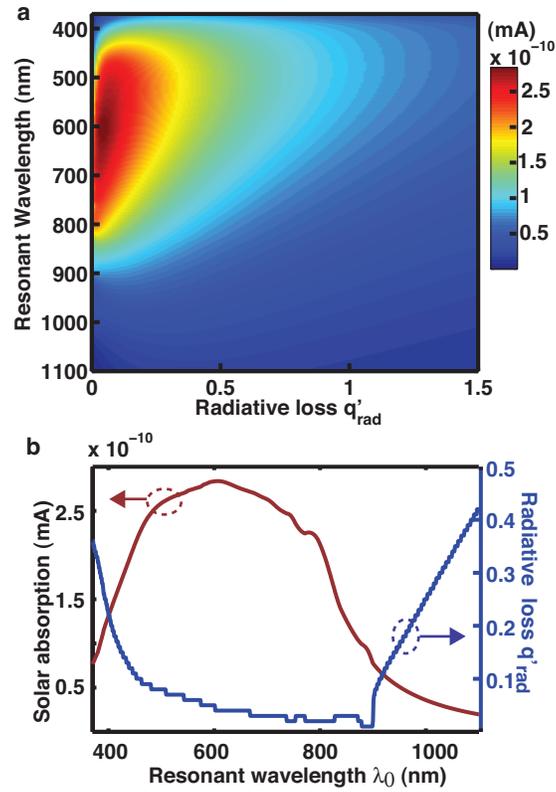

**Figure S2. The solar absorption of single leaky modes in 0D Si structures.** (a) Calculated solar absorption of single leaky modes in 0D Si structures as a function of radiative loss (horizontal axis) and resonant wavelength (vertical axis). (b) The optimal absorption (red line) and associated radiative loss (blue line) as a function of the resonant wavelength.

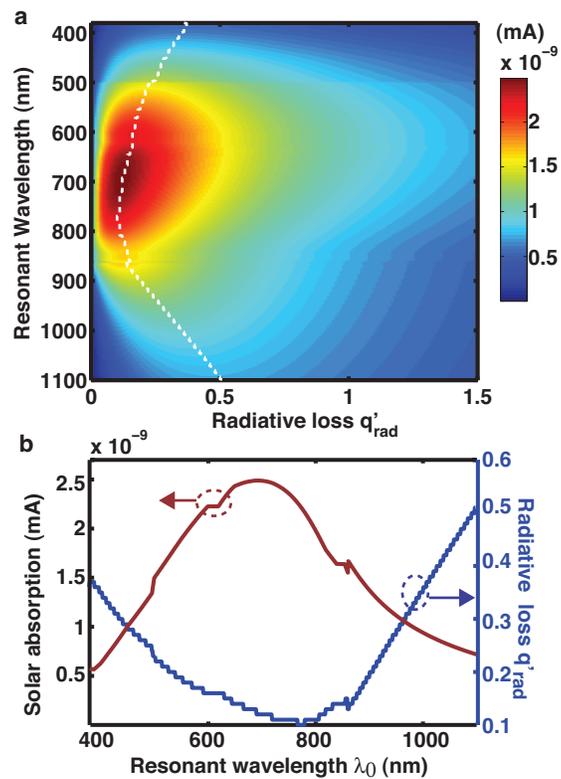

**Figure S3. The solar absorption of single leaky modes in 0D CdTe structures.** (a) Calculated solar absorption of single leaky modes in 0D CdTe structures as a function of radiative loss (horizontal axis) and resonant wavelength (vertical axis). (b) The optimal absorption (red line) and associated radiative loss (blue line) as a function of the resonant wavelength.

## S1. Calculation for the density of leaky modes

For the convenience of mode identification, most of this modal analysis is performed with 1D structures, in which the leaky mode can be identified easier. However, the principle obtained from 1D structures can apply to 0D structures as well.

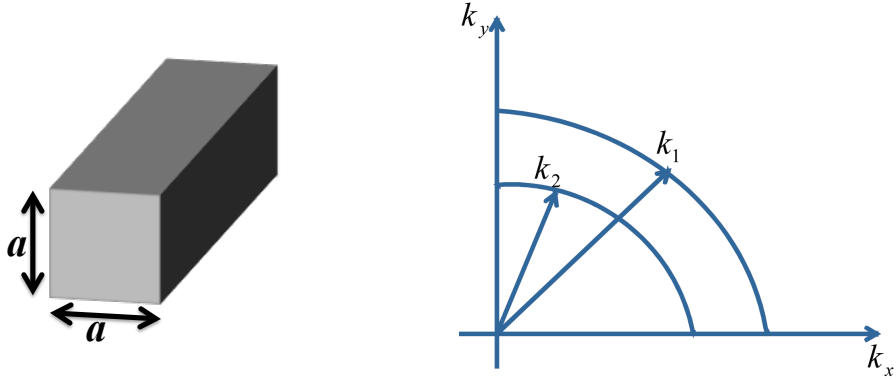

Figure S4. Schematic illustration for 1D square nanowires (left) and for its corresponding mode distribution in k space (right).

we first examine 1D square structure with a refractive index of $n$ as illustrated in Figure S4. We consider modes in wave vector $k$ space. The interspace between neighboring modes in x direction and y direction can be written as $\Delta k_x = \pi/na$ and $\Delta k_y = \pi/na$. The space occupied by one mode is space $\Delta k_x \Delta k_y$. For a given space between $k_1$ and $k_2$, $1/4 \cdot \pi(k_1^2 - k_2^2)$, the number of modes can be written as

$$N(\lambda_1, \lambda_2) = \frac{1}{4} \frac{\pi(k_1^2 - k_2^2)}{\Delta k_x \Delta k_y} = \pi \cdot a^2 \cdot n^2 \cdot (\frac{1}{\lambda_1^2} - \frac{1}{\lambda_2^2}) = \pi A n^2 (\frac{1}{\lambda_1^2} - \frac{1}{\lambda_2^2}) \quad (S1)$$

Similarly, we also derive the number of modes in 0D cubic structure as

$$N(\lambda_1, \lambda_2) = \frac{1}{8} \frac{4}{3} \frac{\pi(k_1^3 - k_2^3)}{\Delta k_x \Delta k_y \Delta k_z} = \frac{4\pi}{3} \cdot a^3 \cdot n^3 \cdot (\frac{1}{\lambda_1^3} - \frac{1}{\lambda_2^3}) = \frac{4\pi}{3} V n^3 (\frac{1}{\lambda_1^3} - \frac{1}{\lambda_2^3}) \quad (S2)$$

From eqs. (S1)-(S2), we can find out the density of modes at given wavelength $\lambda_0$ as

$$\text{for 1D,} \quad \rho(\lambda_0) = \frac{dN}{d\lambda} = \frac{2\pi A n^2}{\lambda_0^3} \quad (S3)$$

$$\text{for 0D,} \quad \rho(\lambda_0) = \frac{dN}{d\lambda} = \frac{4\pi V n^3}{\lambda_0^4} \quad (S4)$$

For the case of solar light absorption, we need take into account both transver magenetic (TM) and transverse electric (TE) modes. Therefore, the density of leaky modes for two polarization are:

for 1D, $$\rho(\lambda_0) = \frac{4\pi A n^2}{\lambda_0^3} \quad (S5)$$

for 0D, $$\rho(\lambda_0) = \frac{8\pi V n^3}{\lambda_0^4} \quad (S6)$$

While derived from square and cubic structures, these equations can apply to nanostructures with arbitrary shapes, as illustrated in Figure S5.

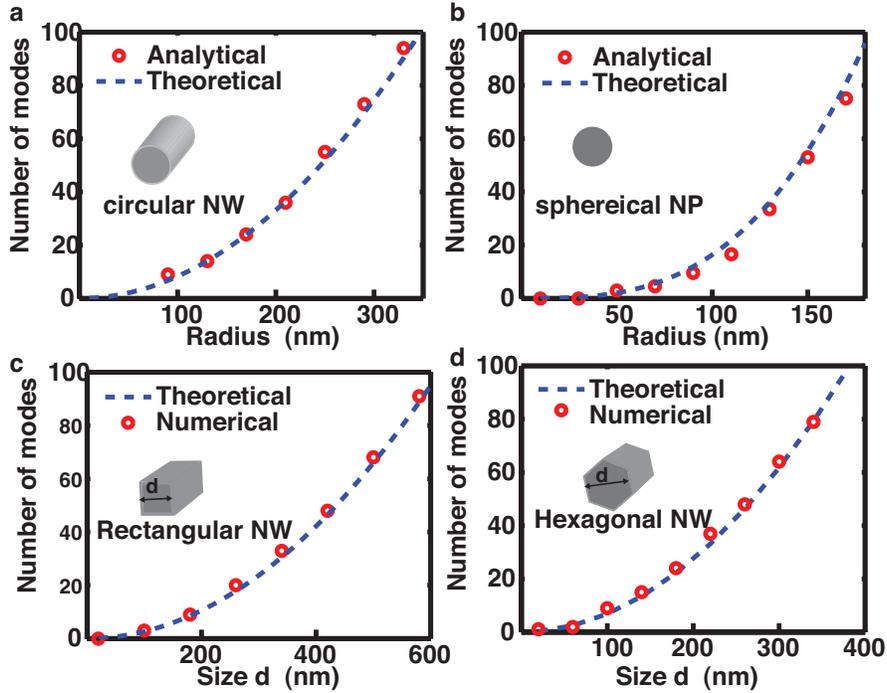

Figure S5. The number of leaky modes calculated using eqs.(S1)-(S2) as a function of size, which is referred as theoretical and plotted as dashed lines. Also plotted is the number of leaky modes directly counted from the eigenvalue calculation of leaky modes using the methods we reported previously.[1-3] We can see that eqs.(S1)-(S2) can indeed reasonably describe the number of leaky modes in nanostructures with arbitrary shapes.

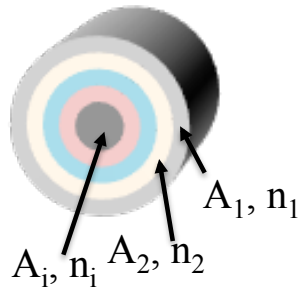

Figure S6. Schematic illustration for heterostructures with absorbing and non-asborbing materials.

We can calculate the number of leaky modes in heterogeneous structures that consist of multiple materials as illustrated Figure S6.

$$\text{for 1D,} \quad N(\lambda_1, \lambda_2) = \pi \cdot (1/\lambda_1^2 - 1/\lambda_2^2) \cdot \sum_i n_i^2 A_i \quad (S7)$$

$$\text{for 0D,} \quad N(\lambda_1, \lambda_2) = \frac{4}{3} \cdot \pi \cdot (1/\lambda_1^3 - 1/\lambda_2^3) \cdot \sum_i n_i^3 V_i \quad (S8)$$

The subscript $i$ here denote different parts of the structure. $n_i$ is the refractive index of the $i$th part, and $A_i$, $V_i$ is the corresponding area or volume of the $i$th part. The validity of these equations is confirmed in Figure S7a, which shows a good consistence between the number of leaky modes calculated using eq. S7 and that counted from eigenvalue calculation[1-3]. We believe that only those leaky modes mainly related with the absorbing materials may substantially contribute to light absorption.

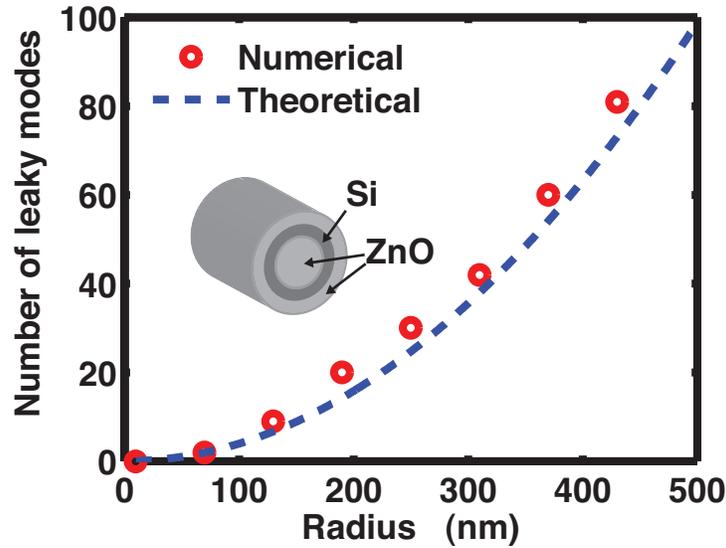

Figure S7. The number of leaky modes for a ZnO/Si/ZnO nanowire, with a constant size ratio of 0. 24, 0.36, and 0.4 from inner to outside, calculated using eqs.(S7) as function of the radius, referred as theoretical and plotted as dashed blue line. Also plotted is the number of leaky modes directly counted from the eigenvalue calculation of leaky modes using COMSOL, referred as numerical and plotted as red circle. Inset is a schematic illustration of the NW.

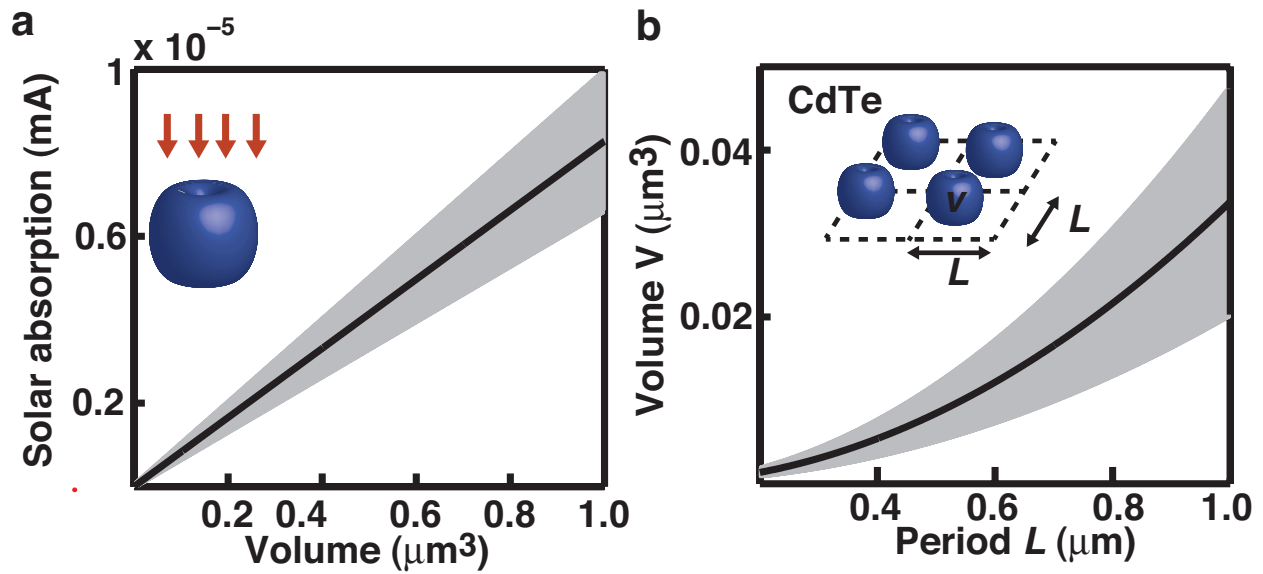

**Figure S8. Solar superabsorption in single nanostructures and an array of nanostructures.** (a) The maximal solar absorption of single 0D nanostructures in which CdTe is included as the absorbing materials as a function of the volume of CdTe materials. The calculation result includes an estimated 20% error as indicated by the shaded areas. The inset is a schematic illustration for the nanostructure, whose irregular shape is intentionally chosen to illustrate that the structure may have any arbitrary shapes. (b) Solar superabsorption limit of an array of 0D nanostructures. The minimum volume of CdTe materials necessary to absorb >90% of the solar radiation above the band gap is plotted as a function of the period of the array. The inset is a schematic illustration for the nanostructure array.

**Table S1. Calcuated eigenvalue and radiative loss for the leaky modes in different structures**

|  | Structure 1 | | Structure 2 | | Structure 3 | |
|---|---|---|---|---|---|---|
| Mode | Eigen value (ka) | Radiative loss ($q'_{rad}$) | Eigen value (ka) | Radiative loss ($q'_{rad}$) | Eigen value (ka) | Radiative loss ($q'_{rad}$) |
| TM21 | 0.93 - 0.019i | 0.014 | 1.40 - 0.098i | 0.07 | 1.46 - 0.225i | 0.154 |
| TM31 | 1.26 - 0.003i | 0.003 | 1.85 - 0.044i | 0.023 | 1.72 - 0.187i | 0.109 |
| TM41 | 1.57 - 0.0007i | 0.0004 | 2.28 - 0.019i | 0.008 | 2.71 - 0.179i | 0.066 |
| TM61 | 2.18 - 0.00003i | 0.00001 | 3.07 - 0.003i | 0.0008 | 2.51 - 0.054i | 0.023 |
| TE21 | 1.23 - 0.017i | 0.013 | 2.06 – 0.145i | 0.07 | 1.83 - 0.176i | 0.096 |
| TE31 | 1.55 - 0.003i | 0.002 | 2.58 - 0.094i | 0.036 | 2.08 - 0.240i | 0.115 |

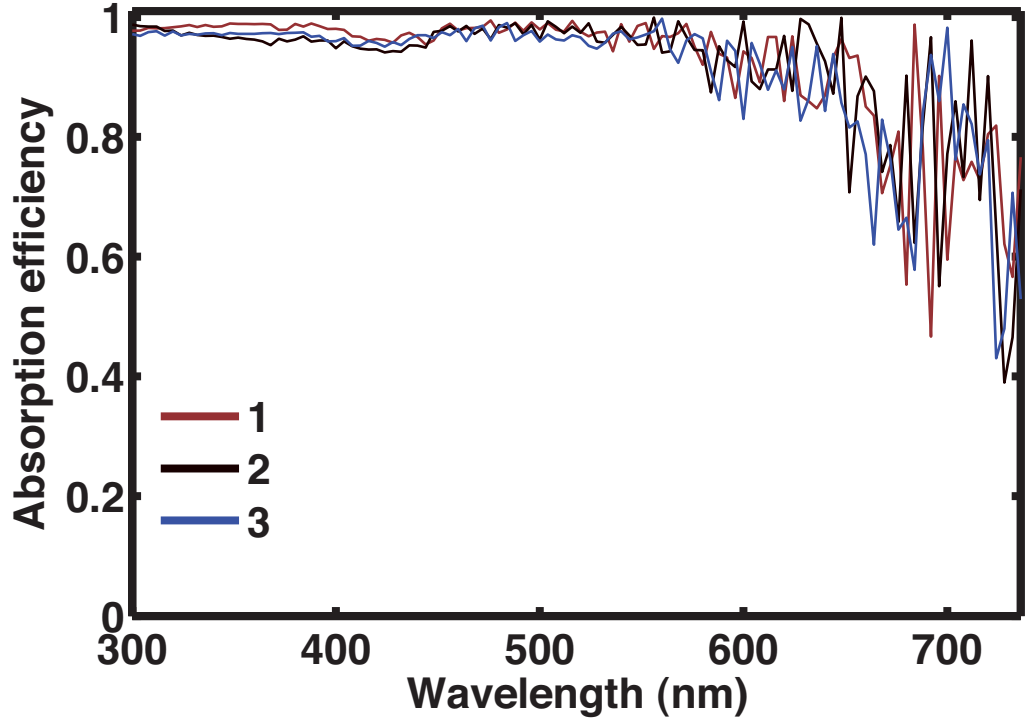

**Figure S9. The robustness of the design.** The calculated spectral absorption efficiency for the designed nanosstructure arrays with a little bit difference in geometrical features. The structure 1 is the one studied in the main text. The detailed geometrical features of the three structures are listed in the following table.

| Structure | side | height | a-Si | SiC | ZnO | SiO2 | Period | Efficiency |
|---|---|---|---|---|---|---|---|---|
| 1. | 180nm | 380nm | 10nm | 30nm | 30nm | 50nm | 540nm | 20.69mA/cm$^2$ 90.8% |
| 2. | 190nm | 390nm | 10nm | 20nm | 40nm | 40nm | 540nm | 20.58mA/cm$^2$ 90.3% |
| 3. | 170nm | 370nm | 10nm | 40nm | 20nm | 20nm | 480nm | 20.35mA/cm$^2$ 89.3% |

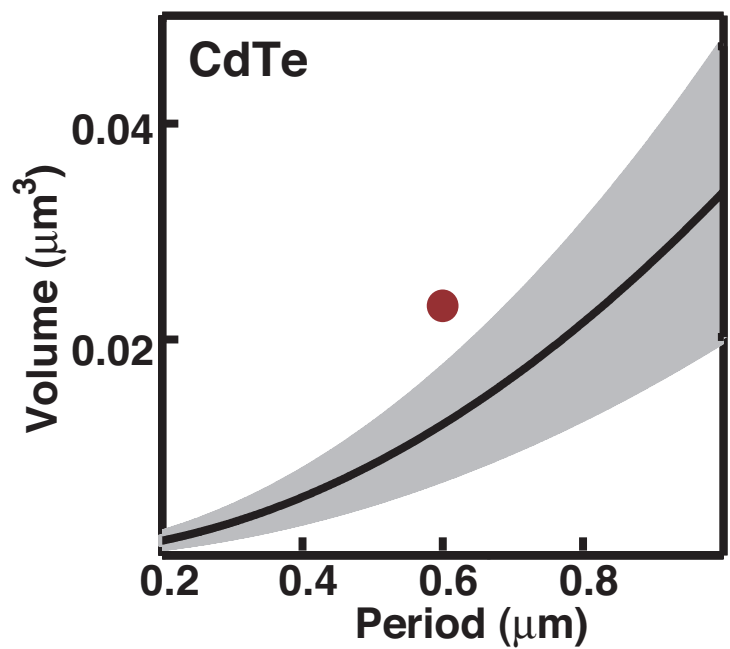

**Figure S10.** The relationship between the designed structure (the red dot) and the predicted minimum volume for CdTe. The structure is shown in Figure 5d inset in the main text.

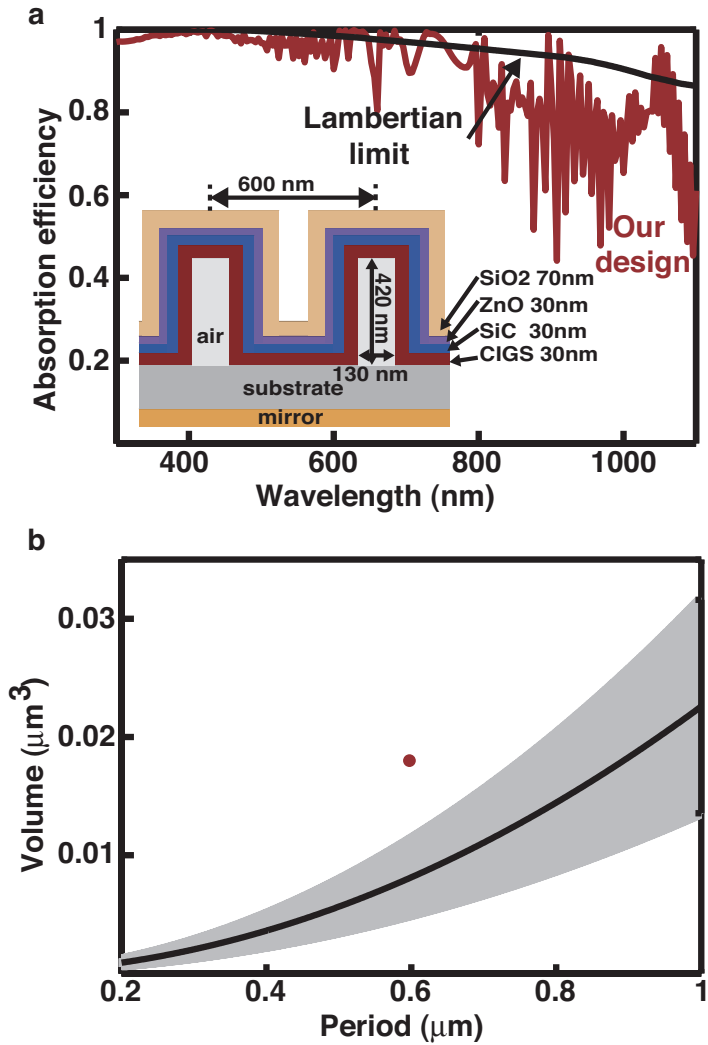

**Figure S11. Design of solar superabsorbers for CIGS.** (a) The calculated spectral absorption efficiency of a designed structure including 30nm thick CIGS. The Lambertian limit for CIGS with an efficient thickness of 52.5 nm is also given (black). The inset shows the geometry of the designed structure. (b) The relationship between the designed structure (the red dot) and the predicted minimum volume for CIGS.